# Retrieval-based and Fine-tuned LLM Approaches for Industrial Asset Health Monitoring and Decision Support

Seshu K. Damarla, *Member, IEEE,* Xiuli Zhu

*Abstract*— **Industrial plants run many important machines like pumps, turbines, and compressors. While engineers can easily use their own experience to find and fix machine problems, it is very hard to teach this smart reasoning to computer systems. This work studies how well a retrieval-only method and an open-source large language model (LLM) reason over failure-sensor diagnostics using the FailureSensorIQ benchmark (a multiple-choice question-answering task) introduced by IBM Research. In the retrieval-only approach, each answer option is converted into an option-level query and scored using similar correct and incorrect records from the training data. TF-IDF, BM25, semantic search, and hybrid search are tested and compared. In the LLM-based approach, Qwen2.5-7B-Instruct model is evaluated using zero-shot prompting, few-shot prompting, and fine-tuned using QLoRA. The results show that semantic search and hybrid search are stronger than pure keyword matching techniques, indicating that meaning-based similarity is more important for the industrial failure–sensor reasoning task. Among the LLM-based methods, the fine-tuned model gives the best performance, substantially improving over zero-shot and few-shot prompting. Error analysis shows that performance decreases as the number of answer options increases, and robustness analysis shows that all methods are sensitive to option shuffling, changed labels, paraphrasing, and added distractors.**

## I. Introduction

Large language models (LLMs) are massive neural networks trained on vast text datasets and have applications in education, healthcare, customer service, research, programming, and content creation. Recent research reveals that LLMs can address several practical problems across major areas of plant operations. Zhang et al. (2025) reviewed how LLMs can assist in intelligent manufacturing and Industry 5.0 [1], while Bajestani et al. (2026) studied how humans and LLMs can work together to make better decisions in smart factories [2]. Garcia et al. (2024) created a framework connecting an LLM directly to a plant's live sensor data, memory logs, and external digital tools [3]. Similarly, Bianchini et al. (2026) built a voice and image assistant to guide assembly line operators [4].

François et al. (2025) showed that LLMs can help plant workers understand, document, and update low-code automated workflows [5]. Fares (2026) discussed the use of LLMs in writing code for programmable logic controllers (PLCs) [6]. Industries possess process manuals, training documents, historical maintenance logs, work orders, etc. It is a gigantic task for engineers/plant operators to read through them and find relevant and useful information to prepare remedial actions for unwanted events. LLMs can read these technical documents and discover answers much faster. Tizaoui and Tan (2024) built a testing dataset for process automation questions and pointed out that general LLMs need specific factory datasets to be truly useful [7].

To improve accuracy, Liu et al. (2026) created a search-and-answer system that pulls matching facts from technical documents before the LLM answers, which drastically improves final answer accuracy [8]. In control rooms, LLMs currently work best as a backup assistant rather than running the plant directly. Wen et al. (2026) found that an LLM performed much better as a supervisory PID tuner than trying to control a water-tank system directly [9]. Vyas and Mercangöz (2025) created a team of AI agents (an operator, a validator, and a reprompter) that double-check each other's work to keep the plant safe [10]. Additionally, Liang et al. (2026) coupled an LLM with a commercial plant simulator [11], while Lima et al. (2026) tested AI agents for predicting chemical properties and advanced plant control [12]. LLMs are highly effective at interpreting unfamiliar operational patterns and explaining what went wrong when a machine breaks down. Chen and Imani (2025) built *MoXpert*, which uses images and engineering facts to spot factory anomalies [13]. Singh et al. (2026) developed a RAG-based oscillation analysis framework that connects LLMs with oscillation detection tools [14]. Xue et al. (2025) proposed StictionGPT, which converts control-loop time-series data into diagnostic images and uses large vision-language models for valve stiction detection [15].

Industrial plants rely on Failure Modes and Effects Analysis (FMEA) to connect a specific equipment failure to abnormal sensor reading. In industrial asset health monitoring, LLMs can be used as decision-support tools to schedule maintenance works or find the root cause of a process/equipment failure. When a sensor shows abnormal readings, an LLM-based assistant can help interpret the situation by linking the abnormal sensor to possible failure modes, relevant assets, and maintenance actions. For example, if temperature changes abnormally, the assistant can support the operator by suggesting likely faults, explaining the reasoning in natural language. To see if LLMs can handle this, a team at IBM Research (Constantinides et al. (2025)) created

Seshu K. Damarla is with Department of Chemical Engineering, University of Alberta, Edmonton, Alberta, T6G2R3 Canada (corresponding author's e-mail: damarla@ualberta.ca).

Xiuli Zhu is with Department of Optical-Electrical and Computer Engineering, University of Shanghai for Science and Technology, Shanghai, P. R. China (e-mail: xiulizhu@usst.edu.cn).

FailureSensorIQ benchmark dataset (a multiple-choice question-answer task) [16]. The team discovered that large and small LLMs performed poorly on the benchmark dataset. This poor performance can be attributed to their training on general text. So, they do not inherently understand specialized industrial machines, how different sensors interact, or how a specific chemical process behaves. Retrieval-augmented generation can provide domain knowledge, but directly adding retrieved text to the prompt may introduce irrelevant or conflicting information and confuse the model. Fine-tuned models can improve performance on standard test data, but their reliability must be checked under practical changes such as question paraphrasing, option shuffling, changed option labels, and stronger distractor options. These issues are especially important for industrial failure-sensor reasoning, where a wrong answer may mislead maintenance or operator decision-making.

Motivated by these gaps, the present work addresses industrial machine-and-sensor reasoning by testing two completely different approaches: retrieval-only methods without LLMs and LLM-based methods. The main contributions of the present work are summarized below:

- Instead of relying on prompt-based RAG, we developed an option-wise retrieval methodology, where each answer option is scored separately using similar evidence from the training data.
- We compared different search tools (TF-IDF, BM25, semantic search and hybrid search) to see which finds the best engineering evidence.
- We first evaluated the capability of a general LLM (Qwen2.5-7B-Instruct), with zero-shot and few-shot prompting, to answer operator's questions. Then, LLM was fine-tuned using the parameter-efficient fine-tuning technique: QLoRA, and subsequently tested on an independent test dataset.
- We evaluated the robustness of the pre-trained and fine-tuned LLM on two perturbed test datasets (options are shuffled, questions are rewritten, and LLMs are given multiple correct answers).
- We provided a deep breakdown of exactly where the proposed methods make mistakes, sorted by number of questions, question type, and whether the search tool and the LLM disagreed.

## II. BACKGROUND AND PROBLEM FORMULATION

Industrial plants rely on field assets (pumps, turbines, compressors, boilers, separation columns, etc.) for processing raw materials into finished products while making sure safety and reliability. Hardware sensors are utilized to measure operational variables (temperature, pressure, flowrate, and so on) for monitoring health status of the assets. One or more variables behave abnormally in the event of an asset failure. Due to this, maintenance engineers/plant operators are required to answer failure mode to sensor questions (if a particular failure occurs, which sensor readings likely show that failure?) and sensor to failure mode questions (if a sensor shows an unusual reading, which failure modes are likely to be responsible?). This kind of logical thinking is common in asset health monitoring, fault diagnosis and maintenance scheduling. Failure Modes and Effects Analysis (FMEA) is used in practice to curate this knowledge. Figure 1 shows the application of LLMs to answer these questions in place of human domain experts. To investigate this problem, we use the FailureSensorIQ dataset [16]. This dataset is developed to assess if artificial intelligence can explain relationships between field assets, failure modes and sensor readings. Each sample in the dataset is a multiple-choice question-and-answer. The dataset comprises several diagnostic questions for the ten field assets shown in Table I. As Table II shows, all questions in the dataset do not have the same number of answer options. This makes the problem harder for LLMs to pick a right answer when it is provided with more answer choices. Figure 2 displays two sample questions from the FailureSensorIQ dataset. Every question has its own metadata, which helps identify the asset, question type, task direction, number of options and correct answer. The representative samples demonstrate that it is not a trivial problem. LLMs need to grasp the asset type, failure modes, sensor measurements, and relationships between them.

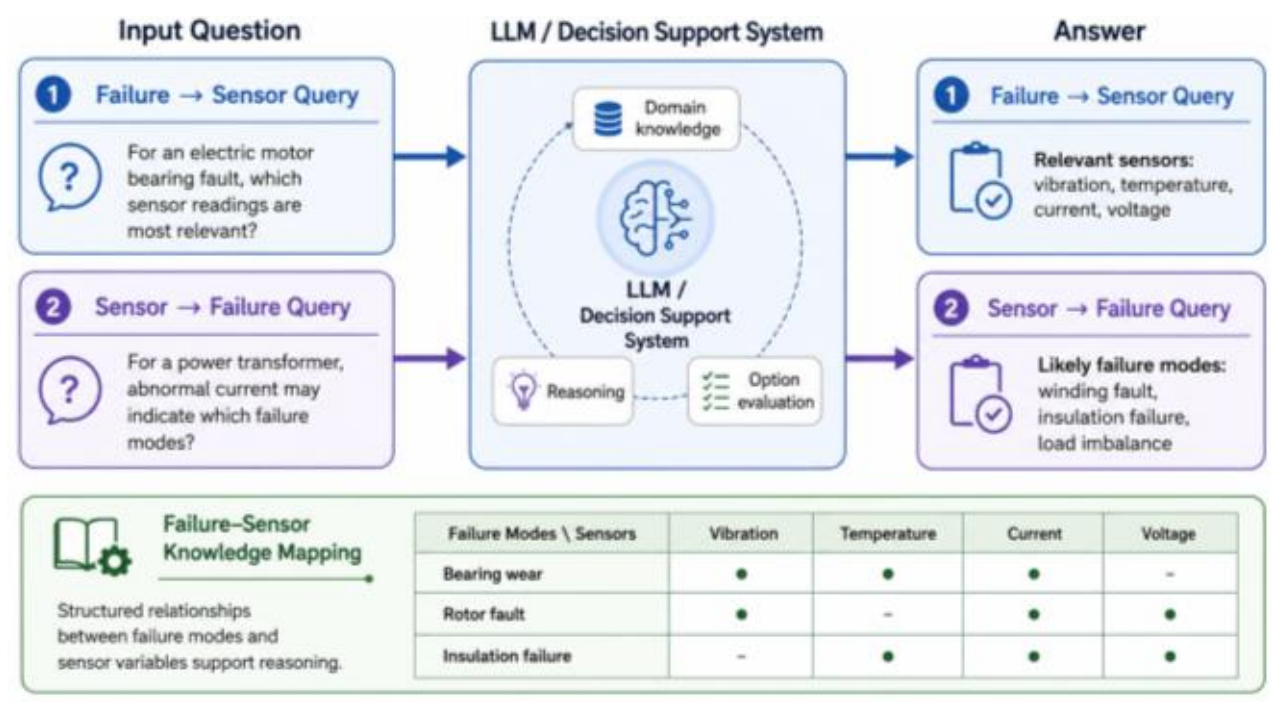

Figure 1. LLM-Based Interpretation of Failure Mode and Sensor Relationships

TABLE I. DISTRIBUTION OF QUESTIONS BY ASSET TYPE

| Asset | No. of Questions |
|---|---|
| Electric motor | 234 |
| Steam turbine | 171 |
| Aero gas turbine | 336 |
| Industrial gas turbine | 240 |
| Pump | 152 |
| Compressor | 220 |
| Reciprocating internal combustion engine | 336 |
| Electric generator | 234 |
| Fan | 200 |
| Power transformer | 544 |
| Total | 2667 |

TABLE II. DISTRIBUTION OF QUESTIONS BY NO. OF ANSWER OPTIONS

| Number of options | Number of questions |
|---|---|
| 2 | 487 |
| 3 | 266 |
| 4 | 389 |
| 5 | 1525 |
| Total | 2667 |

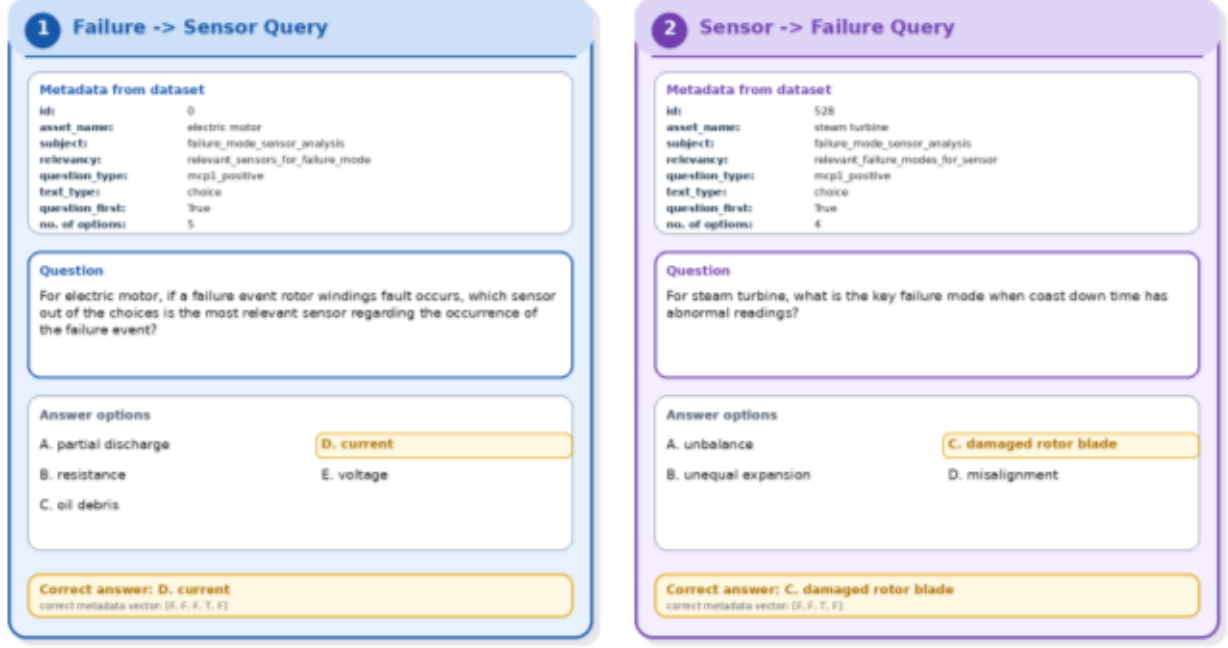

Figure 2. Sample questions from FailureSensorIQ.

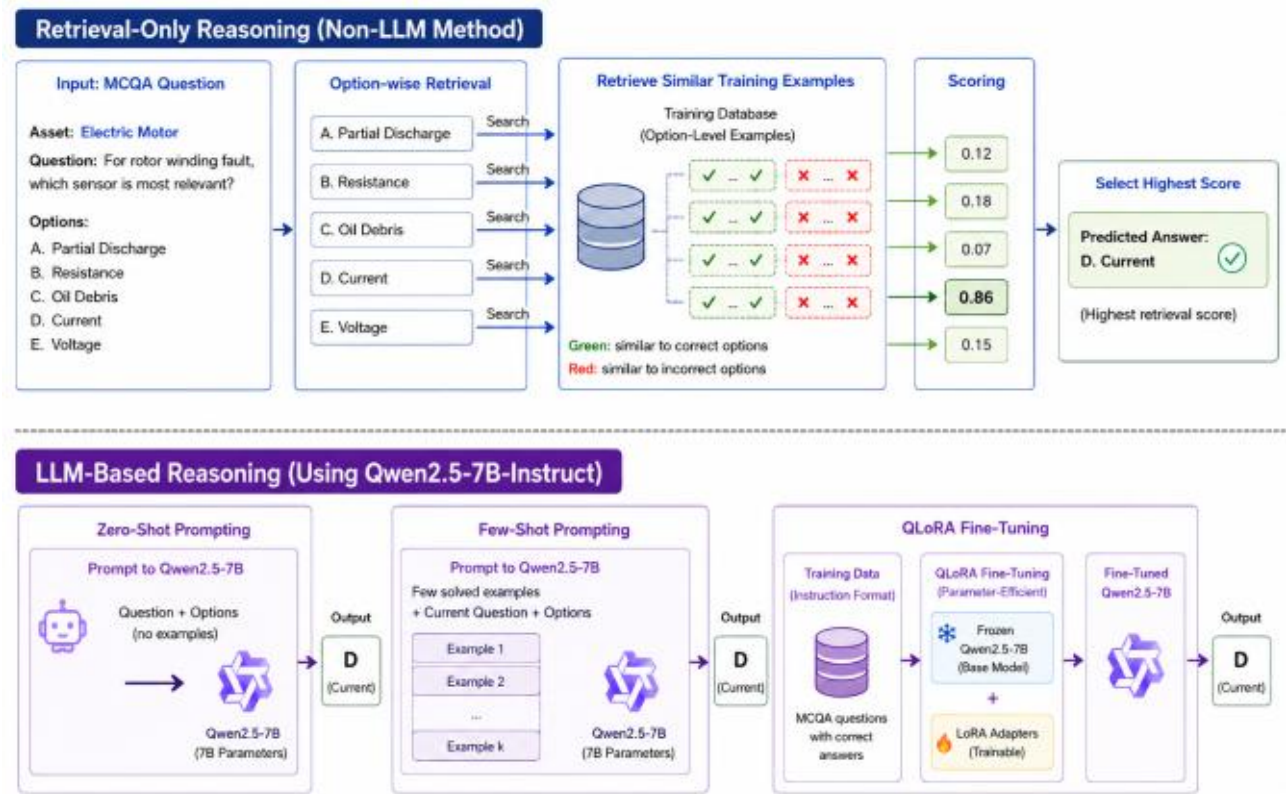

Figure 3. Overview of retrieval-only and LLM-based reasoning methods

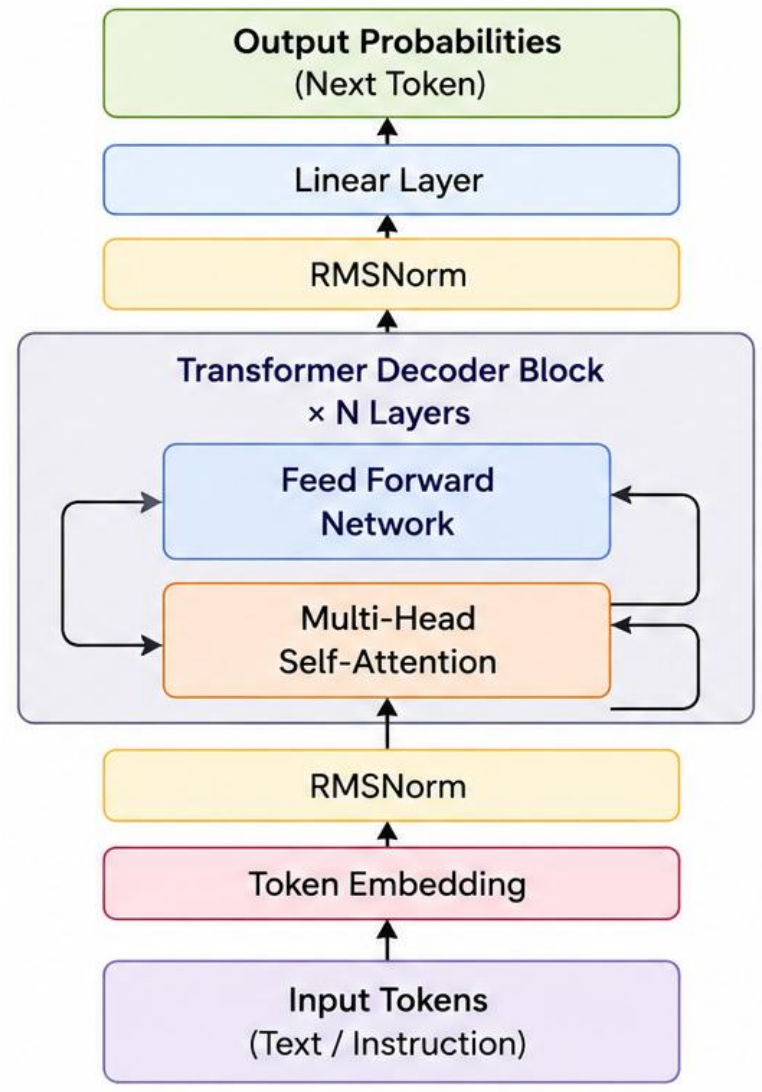

Figure 4. Configuration of Qwen2.5-7B-Instruct model

## III. THE PROPOSED METHODOLOGIES

This section provides detailed description of the methodologies (retrieval-only reasoning and LLM-based reasoning) developed to answer the industrial failure-sensor multiple-choice questions. Figure 3 bestows an overview of the two methods.

### III.A. Retrieval-only reasoning

The retrieval-only reasoning method neither need any LLM nor training. The working principle of the method is quite straightforward. If a test question is akin to questions in the training dataset, then the answer can be chosen by matching the options of the test question with similar questions from the training dataset. For each test question, the method compares each answer option with similar training examples and computes a retrieval score. The option that has the highest retrieval score is picked as the final answer. Since there is no training or fine-tuning involved, the method provides answers very quickly.

In normal retrieval, the full question is used to retrieve similar training examples. This approach may fail to clearly point to the correct answer. Thus, the present work proposes an option-wise retrieval method, where each candidate answer option is verified independently. For instance, consider the following question.

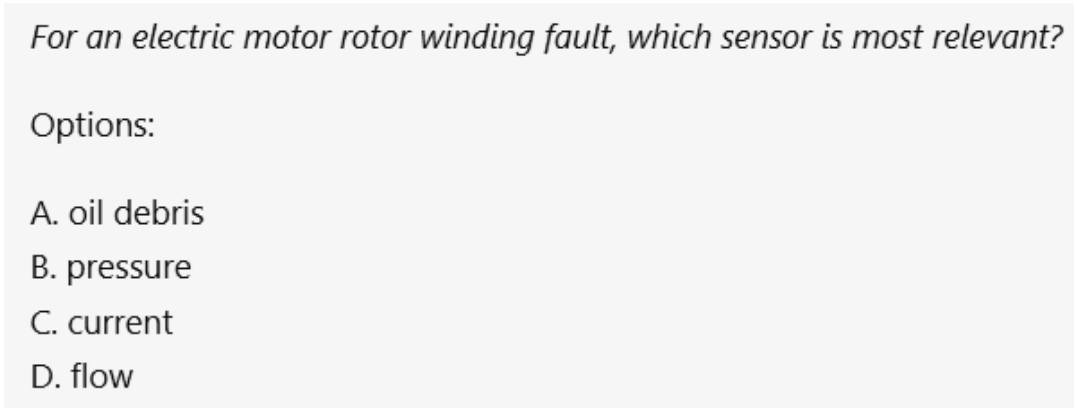
*For an electric motor rotor winding fault, which sensor is most relevant?*

Options:

A. oil debris
B. pressure
C. current
D. flow

Rather than requesting the retrieval system to respond to the complete question, we design individual option-level queries:

- Is **oil debris** relevant to electric motor rotor winding fault?
- Is **pressure** relevant to electric motor rotor winding fault?
- Is **current** relevant to electric motor rotor winding fault?
- Is **flow** relevant to electric motor rotor winding fault?

Each option-level query is judged against option-level records from the training data. The training data is utilized as the retrieval memory. Every training question is changed to option-level records. If a training question has five answer options, then five option-level records are created. The following is an example of what an option-level record holds.

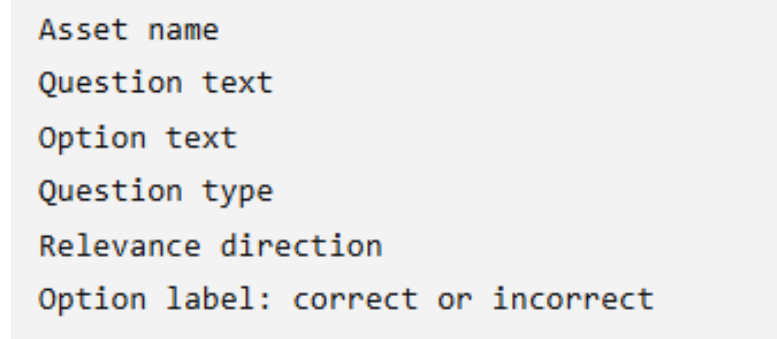
```
Asset name
Question text
Option text
Question type
Relevance direction
Option label: correct or incorrect
```

One option-level record is marked as correct, and the remaining records are marked as incorrect. Because of this conversion, the retrieval system learns the two types of evidence: support records (similar training records where the option is correct) and contradiction record (similar training records where the option is incorrect). Similarity between each option-level record of a test question and each of the training option-level records is measured using keyword search algorithm (TF-IDF or BM25), semantic search algorithm or hybrid search algorithm, which yields top k support and contradict records (in the present work, k assumes 2). The final retrieval score is calculated as given below.

$$retrieval\ score\ = mean(top\ k\ support\ similarities) - \lambda \times mean(top\ k\ contradict\ similarities)$$

where $\lambda = 0.75$.

The retrieval score is computed for every option-level record of the test question and the option having the highest retrievals score is selected as the final answer.

## III.B. LLM-based Reasoning

Qwen2.5-7B-Instruct is an open-source and advanced LLM developed by Alibaba Cloud and the Qwen team [17]. It has 7 billion parameters and is precisely instruction-tuned, making it adept at handling a wide variety of tasks such as reasoning, coding, knowledge retrieval, vision understanding, audio understanding, etc. The configuration of the model is depicted in Figure 4. The model is built based on transformer decoder. There are 28 transformer decoder blocks, 2 RMSnorm layers, a token embedding layer, a linear layer and an output layer. The model was pre-trained using a massive text data and then post-trained using supervised instruction fine-tuning and reinforcement learning to enhance instruction following, reasoning and structured response generation. When the model is given an input, the token embedding layer converts the input text into mathematical vectors that are normalized by the RMSNorm layer to make the input processing stable. The normalized embedding vectors are sent to the series of the transformer decoder blocks (which are backbone of the model). The output of the last decoder block is further processed in the linear layer, which turns the processed vectors back into language, computing the most probable next token to generate a response.

In the present work, the Qwen2.5-7B-Instruct model is evaluated in three setting: zero-shot prompting, few-shot prompting and QLoRA fine-tuning.

***Zero-shot prompting***

In the zero-shot prompting, the model is provided just the test questions and answer options. Correct answer is not provided in the prompt. The model needs to choose a correct option depending on its internal knowledge and reasoning ability. A zero-shot prompt may look like the following.

```
You are an industrial asset health monitoring assistant.
Read the question and select the most correct option.

Question:
For an electric motor, if the failure event is rotor windings fault,
which sensor is the most relevant?

Options:
A. partial discharge
B. resistance
C. oil debris
D. current
E. voltage

Return only the correct option letter.
```

The model output is then verified against the correct answer. The zero-shot prompting assesses if the model can solve the task without fine-tuning or examples.

***Few-shot prompting***

In the few-shot prompting, a few examples (i.e. similar questions with corresponding correct answers) are appended to the test question. These examples teach the model how to answer the test question correctly. A few-shot prompt may be like the following.

```
You are an industrial asset health monitoring assistant.
Select the correct answer option.

Example 1:
Question: ...
Options: ...
Answer: C

Example 2:
Question: ...
Options: ...
Answer: A

Now answer the following question:
Question: ...
Options: ...
Return only the correct option letter.
```

The few-shot prompting is employed as an elementary adaptation method, and it does not modify the weights of the model.

***QLoRA fine-tuning***

As LLMs are pre-trained using a massive corpus of general text, they may not be able to handle domain-specific tasks. As a result, the zero-shot or the few-shot prompting may not yield satisfactory results. So, it is important to adapt the Qwen2.5-7B-Instruct model to the industrial failure-sensor reasoning task. The full fine-tuning of the model needs large memory, computation and huge amount of training data. Therefore, parameter-efficient fine-tuning technique: QLoRA (Quantized Low-Rank Adaptation) is used in this work [18]. In QLoRA, the model weights are frozen, hence, they are not updated. Small trainable layers (LoRA adapters) are added to the model. Only the weights of these LoRA adapters are trained. The Qwen2.5-7B-Instruct model keeps its original knowledge obtained from pre-training whereas the LoRA adapters learn the industrial failure-sensor reasoning task. QLoRA avoids retraining all 7 billion parameters of the Qwen2.5-7B-Instruct model. During fine-tuning, each training sample is transformed into an instruction-answer format shown below.

```
Instruction:
Answer the following industrial failure-sensor multiple-choice question.

Question:
For an electric motor, if the failure event is rotor windings fault,
which sensor is the most relevant?

Options:
A. partial discharge
B. resistance
C. oil debris
D. current
E. voltage

Answer:
D
```

After fine-tuning, the model is appraised on the test set, which helps to understand if domain-specific (or task-specific) fine-tuning handles the task better than the zero-shot prompting and the few-shot prompting.

TABLE III. RESUTLS FOR TEST DATASET

| Method | Acc (%) | NCA | NWA | ITSQ |
|---|---|---|---|---|
| TF-IDF | 56.18 | 300 | 234 | 0.0166 |
| BM25 | 59.18 | 316 | 218 | 0.0231 |
| Semantic search | 70.97 | 379 | 155 | 0.0362 |
| Hybrid search | 73.41 | 392 | 142 | 0.0696 |
| ZS Qwen2.5-7B | 55.43 | 296 | 238 | 0.1931 |
| FS Qwen2.5-7B | 54.49 | 291 | 243 | 0.3983 |
| QL Qwen2.5-7B | 88.01 | 470 | 64 | 0.2549 |

ACC – accuracy, NCA – no. of correct answers, NWA – no. of wrong answers, ITSQ – inference time (seconds per question), ZS – zero-shot, FS – few-shot, QL – QloRA.

TABLE IV. ERROR ANALYSIS BY QUESTION TYPE ON TEST DATASET

| Method | Negative Questions Accuracy (%) | Positive Questions Accuracy (%) |
|---|---|---|
| TF-IDF | 61.94 | 44.25 |
| BM25 | 63.61 | 50.00 |
| Semantic search | 79.17 | 54.02 |
| Hybrid search | 80.28 | 59.20 |
| ZS Qwen2.5-7B | 57.50 | 51.15 |
| FS Qwen2.5-7B | 56.67 | 50.00 |
| QL Qwen2.5-7B | 90.00 | 83.91 |

TABLE V. ERROR ANALYSIS BY NO. OF OPTIONS ON TEST DATASET

| Method | 2 Options (%) | 3 Options (%) | 4 Options (%) | 5 Options (%) |
|---|---|---|---|---|
| TF-IDF | 74.77 | 73.68 | 56.00 | 46.10 |
| BM25 | 78.50 | 77.19 | 65.33 | 47.12 |
| Semantic search | 94.39 | 85.96 | 80.00 | 57.29 |
| Hybrid search | 94.39 | 85.95 | 85.33 | 60.34 |
| ZS Qwen2.5-7B | 86.92 | 70.18 | 53.33 | 41.69 |
| FS Qwen2.5-7B | 87.85 | 61.40 | 60.00 | 39.66 |
| QL Qwen2.5-7B | 99.07 | 94.74 | 89.33 | 82.37 |

TABLE VI. DISAGREEMENT ON TEST DATASET

| Method | BC | BW | FMC-SMW | FMW-SMC |
|---|---|---|---|---|
| TF-IDF vs SN | 264 | 119 | 36 | 115 |
| BM25 vs SN | 264 | 103 | 52 | 115 |
| Hybrid vs SN | 361 | 127 | 28 | 15 |
| FS vs QLoRA | 276 | 49 | 15 | 194 |
| FS vs ZS | 245 | 192 | 46 | 51 |
| QLoRA vs ZS | 279 | 47 | 191 | 17 |

BC – both correct, BW – both wrong, FMC-SMW – first method correct -second method wrong, FMW-SMC – first method wrong-second method correct.

TABLE VII. RESUTLS FOR PERTURBED TEST DATASETS

| Method | Acc (%) | ASP (%) | ACP (%) | ACC (%) |
|---|---|---|---|---|
| TF-IDF | 56.18 | 29.59 | 34.08 | 25.28 |
| BM25 | 59.18 | 27.34 | 34.46 | 24.72 |
| Semantic search | 70.97 | 30.52 | 29.96 | 26.78 |
| Hybrid search | 73.41 | 29.21 | 26.97 | 23.78 |
| ZS Qwen2.5-7B | 55.43 | 15.17 | 14.61 | 12.36 |
| FS Qwen2.5-7B | 54.49 | 19.48 | 18.73 | 14.42 |
| QL Qwen2.5-7B | 88.01 | 43.07 | 40.82 | 39.51 |

ASP – accuracy on simple perturbed dataset, ACP – accuracy on complex perturbed dataset, ACC - Consistency-Based Accuracy on the Complex Perturbed Dataset.

## IV. RESULTS AND DISCUSSIONS

The FailureSensorIQ dataset was randomly split into training and test subsets using an 80:20 ratio. The training subset comprised 2133 questions and the test subset had 534 questions. To evaluate the robustness of the proposed methods, simple and complex perturbed datasets were generated from the test set, which avoided data leakage. For the retrieval-based reasoning, each training question was transformed into option-level records. The TF-IDF search method employed unigram and bigram features with L2 normalization. BM25 used $k_1 = 1.5$ and $b = 0.75$. The all-MiniLM-L6-v2 embedding model was used in the semantic search. The hybrid search method was formed as per the following equation.

$$Hybrid\ search = \alpha \times TF-IDF + (1-\alpha) \times semantic\ search$$

where $\alpha = 0.1$.

To fine-tune the Qwen2.5-7B-Instruct model, the model was loaded in 4-bit NF4 quantized mode. The LoRA adapters were applied to the attention and the feed-forward projection layers. The model was fine-tuned using the following hyperparameters: a LoRA rank of 16, alpha of 32, and dropout of 0.05, trained over a single epoch with a learning rate of $2 \times 10^{-4}$, a batch size of 1, a gradient accumulation of 8, and a maximum sequence length of 768. The LLM-based experiments were carried out on a GPU, and the retrieval-based experiments were conducted on a CPU. All these experiments were carried out on a GPU. The results of the proposed methods on the test set are reported in table III. The accuracy of the semantic search and the hybrid search are higher than that of the TF-IDF and BM25 keyword matching methods, which proves that meaning-based similarity is more useful than exact word matching for the industrial failure-sensor reasoning task. The mediocre performance was shown by the zero-shot and few-shot prompting. The fine-tuned model demonstrated the best overall performance, which underscores domain-specific adaption is necessary for the specialised failure-sensor reasoning task. A deep error analysis was carried out based on question type and the number of options available for each question. Tables 4 and 5 show where the models made mistakes. All the methods did better on negative questions than positive ones. This means it is easier for the model to rule out wrong answers than to pick the right one directly. Accuracy also dropped when we added more answer choices, proving that extra trick options make the task harder. The results of the disagreement analysis are provided in Table VI. The semantic search and the hybrid search corrected several errors made by TF-IDF and BM25, showing that meaning-based retrieval captures useful information that keyword methods may miss. For the LLM methods, the fine-tuned model corrected many errors made by the zero-shot and few-shot prompting, confirming the benefit of fine-tuning. Table VII reports robustness results. All methods suffered performance drops on the simple and complex perturbed datasets. This shows that high accuracy on the original test set does not guarantee robustness to option shuffling, changed labels, added distractors, or paraphrased questions. QLoRA remained the strongest method under perturbation, but its performance still dropped considerably.

## V. Conclusions

In the present work, the retrieval-only and the LLM-based methods were developed for handling the industrial failure-sensor reasoning task. Among the retrieval methods, the hybrid method delivered the highest performance, proving that both the contextual meaning and the exact token matching are critical for the reasoning task. The robustness analysis revealed that the methods experienced performance drops under option shuffling, question paraphrasing, changed labels, and additional distractors. Overall, the fine-tuned Qwen2.5-7B-Instruct model is effective for answering the failure-sensor diagnostic questions. Some practical takeaways from the present work are summarized below:

- General purpose LLMs require domain/task-specific adaptation before deployment in specialized industrial decision support.
- The option-wise retrieval methods are faster, simpler and interpretable whereas the fine-tuned Qwen2.5-7B-Instruct model is more accurate. For actual field decision support, a careful combination of these methods may provide more reliable results.
- The proposed methods are meant to augment, rather than replace, human expertise in plant operations and maintenance.
- The present work does not include human-in-the-loop validation as it falls outside the scope of the work.
- Due to scope constraints, only the Qwen2.5-7B-Instruct model was evaluated in this work. Other small LLMs may yield better performance.